\documentstyle[twocolumn,floats,prb,aps]{revtex}

\begin{document}

\draft    % Show PACS

\wideabs{    %Not to use in preprint mode

\title{Localization by disorder in the infrared conductivity of
       (Y,Pr)Ba$_2$Cu$_3$O$_7$ films}
       
\author{R. P. S. M. Lobo\cite{lobomail}}
\address{Laboratoire de Physique du Solide, 
         Ecole Sup\'erieure de Physique et Chimie 
         Industrielles de la Ville de Paris, CNRS UPR 5, 
         75231 Paris Cedex 5, France}

\author{E. Ya. Sherman}
\address{Institute f\"ur Theoretische Physik --
         Karl-Frazens-Universit\"at, A-8010, Graz, Austria}

\author{D. Racah and Y. Dagan}
\address{Department of Physics and Astronomy -- Tel-Aviv University,
         Ramat-Aviv, Tel-Aviv, 69978 Israel}

\author{N. Bontemps}
\address{Laboratoire de Physique du Solide, 
         Ecole Sup\'erieure de Physique et Chimie 
         Industrielles de la Ville de Paris, CNRS UPR 5, 
         75231 Paris Cedex 5, France}

\date{\today}

\maketitle

\begin{abstract}
  The ab-plane reflectivity of (Y$_{1-x}$Pr$_x$)Ba$_2$Cu$_3$O$_7$
  thin films was measured in the 30--30000 cm$^{-1}$ range for
  samples with $x = 0$ ($T_c = 90$ K), $x = 0.4$ ($T_c = 35$ K)
  and $x = 0.5$ ($T_c = 19$ K) as a function of temperature in the
  normal state. The effective charge density obtained from the
  integrated spectral
  weight decreases with increasing $x$. The variation is consistent
  with the higher dc resistivity for $x = 0.4$, but is one order of
  magnitude smaller than what would be expected for $x = 0.5$. In the
  latter sample, the conductivity is dominated at all temperatures by
  a large localization peak. Its magnitude increases as the temperature
  decreases. We relate this peak to the dc resistivity enhancement. A
  simple localization-by-disorder model accounts for the
  optical conductivity of the $x = 0.5$ sample.
\end{abstract}

\pacs{72.15.Rn, 74.25.Gz, 74.76.Bz}

}   %Not to use in preprint mode

\section{Introduction}

The only consensus on the electronic properties of the normal
state of high-$T_c$ superconductors is that they are not
conventional. Examples of models describing cuprates normal state
are charge stripes,\cite{R01} polarons, \cite{R02,R03} various
flavors of modified Fermi liquid \cite{R04,R05,R06} and Luttinger
liquid.\cite{R07} All these models assume a strong
electron-electron and/or electron-phonon interaction. The disorder
introduced by cationic or oxygen doping influences the spectrum 
of excitations making the physics of the system more complex. 
Indeed, Basov {\em et al.} \cite{R10} propose
that the Drude-like peak observed in the infrared spectra of
irradiated YBa$_2$Cu$_3$O$_7$ (YBCO) moves up in frequency to a
localized state. Problems that have been discussed now for
many years are the relevance of disorder to localizazion and 
localization to superconductivity, in particular near the 
metal-insulator transition.\cite{R13,R14,R14b} In cuprates,
this could apply to a strongly underdoped material. Another twist 
appears with the observation of a normal state gap in underdoped 
cuprates.\cite{R08,R09,R11,R12} It was first observed as a 
pseudo-gap in NMR data.\cite{R12b} It was later associated to the 
departure from linearity in the resistivity.\cite{R11,R12} Puchkov 
and coworkers \cite{R08} proposed that the apparent optical 
conductivity spectral weight loss in the normal state of underdoped 
cuprates is a manifestation of the pseudo-gap.

There is a significant (and somewhat contradictory) body of
literature suggesting both hole depletion in the CuO$_2$ planes
and localization in Pr substituted 
compounds.\cite{R18,R19,R20,R21,R22,R23,R24,R25,R25b} 
In this paper we describe
the optical reflectivity of Pr doped YBa$_2$Cu$_3$O$_7$ (Pr-YBCO)
ab-plane thin films as a function of temperature.

The optical conductivity of untwined single crystals of
non-superconducting PrBa$_2$Cu$_3$O$_7$ shows that substitution of
Y by Pr empties the CuO$_2$ planes and localizes the charges into
a mid-infrared band presumably along the CuO chains.\cite{R15}
Optical\cite{R16} ($x \leq 0.35$) and coherent THz\cite{R17} 
($x \leq 0.4$) data on the in-plane response of Pr-YBCO support 
a picture in which the localization grows continuously with Pr 
doping. More recently, Bernhard and co-workers measured the 
far-infrared c-axis conductivity of Pr-YBCO using ellipsometry. 
Their data suggest that Pr acts as an underdoping agent.\cite{R17b}
For their highest doping ($x = 0.45$) they find some indirect 
support of charge localization. However a localization peak has 
not been resolved in Pr-YBCO.

We investigate Y$_{1-x}$Pr$_x$Ba$_2$Cu$_3$O$_7$ samples with $x =
0$, 0.4 and 0.5. The superconducting state of the first two
compositions is discussed in detail elsewhere.\cite{R26} In this
paper we resolve, for the first time, the localization peak in
Pr-YBCO at the composition $x = 0.5$. A simple localization model
based on Pr disorder reproduces this localization peak. We suggest
that a signature of this peak is already present in the $x = 0.4$
compound and competes with the normal state gap opening.

\section{Experimental}

\begin{figure}
\begin{center}
  \input epsf
  \epsfclipon
  \epsfxsize=8cm
  \epsffile{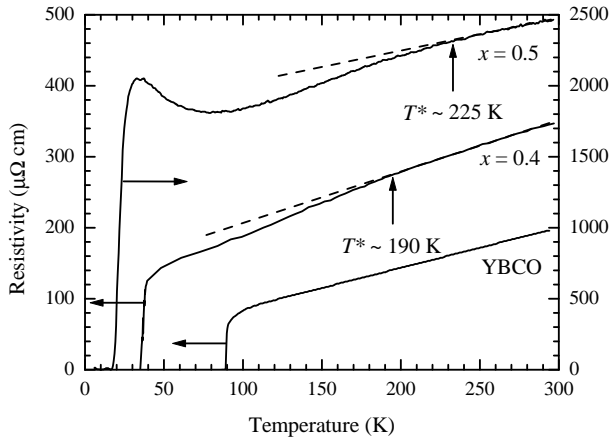}
\end{center}
\caption{Temperature dependence of the dc resistivity of
         (Pr,Y)Ba$_2$Cu$_3$O$_7$ films on YSZ. Note that the
         $x = 0.5$ compound (right-hand axis) is almost one
         order of magnitude more resistive than $x = 0.4$ and
         pure YBCO. The temperature where the resistivity no
         longer shows a linear thermal dependence is indicated
         by $T^*$.}
\label{fig1}
\end{figure}

Thin films of pure and Pr doped YBa$_2$Cu$_3$O$_7$ were grown by
sputtering on stabilized zirconia (YSZ) substrates. The samples
are c-axis oriented and typically 5000 {\AA} thick. The four-point
electrical dc resistivity is shown in Fig. \ref{fig1}, giving $T_c
= 90$K ($x = 0$), $T_c = 35$K ($x =0.4$) and $T_c = 19$K ($x =
0.5$). The resistivity of the $x = 0.5$ compound is almost an
order of magnitude higher (right-hand side scale) than that of
pure or $x = 0.4$ Pr YBCO (left-hand side scale). The dashed
straight lines help to locate approximately the pseudo-gap opening
temperature ($T^*$) in the Pr doped samples, indicated by the
arrows. Note that in the $x = 0.5$ compound an upturn before the
superconducting transition occurs. One relevant point to this work
is that similar $T^*$'s can be seen in underdoped YBCO films
\cite{R11} having $T_c$'s of 70 K and 60 K.

\begin{table}[htb]
  \begin{center}
  \begin{tabular}{ccccc}
    $x$ & $T_c(K)$ & $\Omega_p($cm$^{-1})$ & 
    $N_{eff}($cm$^{-3}$) & $\alpha $($\mu\Omega$cm/K)\\
    \tableline
    0 & 89(1) & 22000 & $5.1 \times 10^{21}$ & 0.57 \\
    0.4 & 35(2.5) & 19200 & $4.1 \times 10^{21}$ & 0.72 \\
    0.5 & 19(7) & 18000 & $3.6 \times 10^{21}$ & 2.3 \\
  \end{tabular}
  \end{center}
  \caption{Pr-YBCO characteristic parameters. $T_c$ and
           $\Delta T_c$ (in parenthesis) are directly
           obtained from the resistivity in Fig. \ref{fig1}.
           $\Omega_p$ and $N_{eff}$ are obtained with
           Eq. \ref{eq3} integrating the room temperature optical
           conductivity up to 1.5 eV. $N_{eff}$ uses the bare
           electron mass as $m^*$. $\alpha$ is the slope of the
           linear part of the resistivity (dashed lines in
           Fig. \ref{fig1}).}
  \label{tab1}
\end{table}

Our infrared reflectivity spectra were obtained with a Bruker IFS
66v interferometer between 30 and 7000 cm$^{-1}$. Near-infrared
and visible data between 4000 and 30000 cm$^{-1}$ were measured in
a Cary 4000 grating spectrometer. In the overlapping spectral
range, measurements from both spectrometers agree within 0.5 \%.
We used gold mirrors as a reference for measures in the Bruker
spectrometer and silver mirrors in the Cary. We utilized a He gas
flow cryostat to measure the spectra between 6 K and room
temperature in the whole frequency range. The temperature
stability during the measurements was better than 0.5 K.

\section{Data Analysis}

The spectral functions were determined for our samples by
Kramers-Kronig transform. At low frequencies (below 30 cm$^{-1}$)
a Hagen-Rubens extrapolation was used at all temperatures,
including the superconducting state. Above the highest measured
frequency (30000 cm$^{-1}$) the reflectivity was assumed to be
constant up to $10^6$ cm$^{-1}$ followed by a $\omega^{-4}$
termination to infinity. Below 150 K our samples are opaque
enough to avoid a significant contribution from the substrate.
We also checked that different low frequency extrapolations
do not change the conductivities more than 7 \% at 100 cm$^{-1}$
and not more than 2 \% at 200 cm$^{-1}$. Above 250 cm$^{-1}$
the differences are negligible in the data of Fig. \ref{fig1}.

A generalization of the Drude model can be obtained using a 
frequency dependent scattering rate\cite{R28,R29} defined as
\begin{equation}
  \frac{1}{\tau} =
     \frac{2\pi}{Z_0}\Omega^2_p\frac{\sigma_1}{\sigma^2_1 + \sigma^2_2},
  \label{eq2}
\end{equation}
$Z_0 = 377 \Omega$ being the vacuum impedance. $\Omega_p$, the
effective plasma frequency, is related to the charge density $n$,
carrier effective mass $m^*$ and the vacuum permittivity $\epsilon_0$ by 
$\Omega_p^2 = ne^2/\epsilon_0 m^*$.\cite{NOTEONPLASMA} 
$\sigma(\omega) = \sigma_1 + i \sigma_2$ is the complex conductivity.

One can estimate the plasma frequency through the classical sum
rule
\begin{equation}
  \Omega^2_p =
     \frac{Z_0}{\pi^2} \int_0^\infty \sigma_1(\omega)d\omega.
  \label{eq3}
\end{equation}
One usually introduces a cutoff energy (1--2 eV) in order to
restrict the integration to the free carrier contribution. 
Eqs. \ref{eq2} and \ref{eq3} assume that $\Omega_p$, $\tau^{-1}$
and $\omega$ are measured in cm$^{-1}$ and $\sigma_1$ in
$\Omega^{-1}{\rm cm}^{-1}$.

\section{Results}

\begin{figure}
\begin{center}
  \input epsf
  \epsfclipon
  \epsfxsize=8cm
  \epsffile{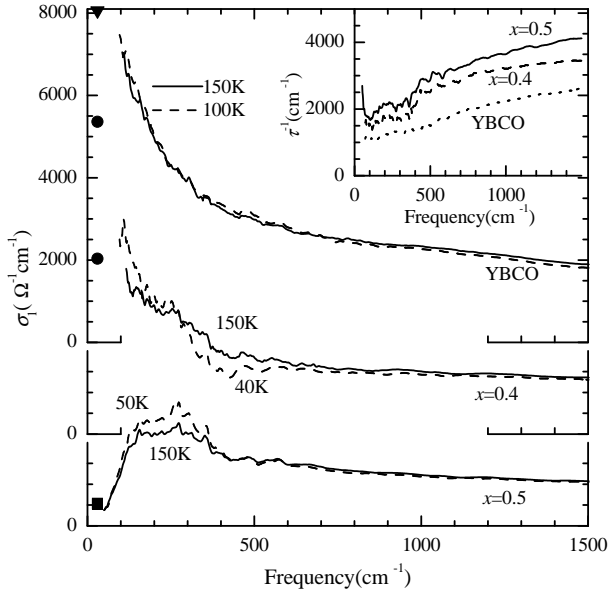}
\end{center}
\caption{Real part of the optical conductivity of Pr-YBCO
         films on YSZ. From top to bottom $x = 0$, 0.4
         and 0.5.
         The solid symbols are obtained from the dc resistivity
         for the corresponding temperatures in the normal state.
         In the $x = 0.5$ sample $\sigma_1$ is dominated by a
         peak around 250 cm$^{-1}$. The inset shows the scattering
         rate at 100 K for all samples. The scales for the $x = 0.4$
         and 0 conductivities are shifted with respect to the x = 0.5 sample 
         by 2000 and 4000 $\Omega^{-1}$cm$^{-1}$, respectively.}
\label{fig2}
\end{figure}

Fig. \ref{fig2} shows the real part of the optical conductivity
for all three samples: $x = 0$, 0.4 and 0.5. The normal state low
frequency conductivity extrapolates consistently at various
temperatures to the measured dc conductivity (solid symbols).

The YBCO film exhibits a conventional behavior. Its spectral
response does not change significantly in the normal state from
150 K to 100 K and is dominated by a Drude-like peak. YSZ is not
the best substrate to grow YBCO. For instance, the residual
conductivity in the superconducting state is twice as high as that
from the best films.\cite{R26} However, our YBCO's normal state
conductivity and scattering rate are qualitatively and
quantitatively the same as those from good samples.\cite{R08}

In the $x = 0.4$ film the free carrier response is still the major
contribution to the normal state conductivity. The loss of
spectral weight from $\sim 800$ cm$^{-1}$ down to 300 cm$^{-1}$
observed at temperatures above $T_c$ is assigned to the opening of
the normal state gap.\cite{R26} It is worth remarking at this
point that a weak shoulder appears at 250 cm$^{-1}$ in the 40 K
spectrum.

A major change, consistent with the huge increase in the dc
resistivity, is observed in the $x = 0.5$ sample. The free carrier
contribution is not seen, and is replaced by a broad peak centered
around 250 cm$^{-1}$ that dominates the far-infrared spectrum
(below 700 cm$^{-1}$). These spectra do not show any signature of
a pseudo-gap or of the superconducting transition. Actually, an
opposite effect is observed since the magnitude of the peak around
250 cm$^{-1}$ increases with decreasing temperature.

In the inset we show the frequency dependent scattering rate
calculated with Eq. \ref{eq2}. The YBCO film displays a standard
behavior with a scattering rate $1/\tau$ depending roughly
linearly on frequency in the normal state. In the Pr substituted
samples the absolute value of $1/\tau$ increases and some
structures appear at low frequencies (250 cm$^{-1}$). Both effects
indicate a lower mobility of charge carriers with a possible
localization effect.

\begin{figure}
\begin{center}
  \input epsf
  \epsfclipon
  \epsfxsize=8cm
  \epsffile{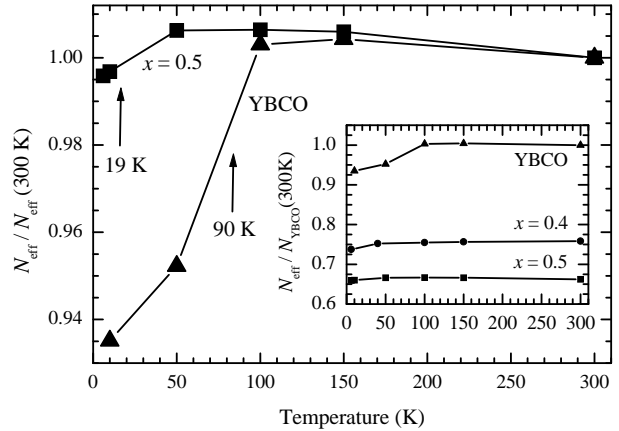}
\end{center}
\caption{Thermal dependence of the charge density (normalized to
         its value at 300 K) calculated with Eq. \ref{eq3}
         integrated up to 1.5 eV. The arrows indicate the critical
         temperature for each sample. The estimated error in
         $N_{eff}$ is roughly twice the symbol size. The inset shows
         the same data, now normalized for all samples to the
         effective charge of YBCO at 300 K.}
\label{fig3}
\end{figure}

The values for $\Omega_p$ estimated from the conductivity
integrated up to 1.5 eV at $T = 300$ K are shown in Table
\ref{tab1}. The charge carrier concentration $N_{eff}$ is derived
from $\Omega_p$ assuming that $m^*$ is the bare electron mass.
In Fig. \ref{fig3}, we display the thermal evolution of $N_{eff}$
for each sample normalized by its value at 300 K. The estimated
error in $N_{eff}$ is about twice the symbol size. The arrows
indicate the measured $T_c$. All samples exhibit a decrease of the
carrier density when crossing $T_c$, associated to the spectral
weight transferred to the zero-centered $\delta$-function. The
inset shows the same data normalized by $N_{eff}$ of YBCO at 300 K
for all samples.

\section{Discussion}

Even though there are not many data points in Fig. \ref{fig3}, we
can infer some general trends for each sample. In pure YBCO we
observe that, within experimental error, $N_{eff}$ does not vary
in the normal state. In the $x = 0.5$ sample $N_{eff}$ increases
when cooling down to 50 K and then begins decreasing. The increase
is consistent with the accumulating spectral weight in the low
frequency peak. The decrease at low temperatures is the signature
of the superconducting transition.

The $x = 0.5$ Pr sample exhibits a clear peak at $\sim 250$
cm$^{-1}$. Experiments in YBCO and pure PBCO have shown a peak in
the conductivity when light is polarized along the b 
axis.\cite{R15,R31} In pure PBCO, the peak at an energy $\sim 1700$
cm$^{-1}$ has been assigned to localized carriers in the 
chains.\cite{R15} In single domain YBCO crystals, the b-axis
conductivity displays a peak at $\sim 300$ cm$^{-1}$, again 
assigned to the chains.\cite{R31} Other authors did suggest disorder
on Pr sites or in oxygen environment.\cite{R22,R32} 2D
localization (hence within the CuO$_2$ planes) has also been
proposed.\cite{R24,R32} Localization by disorder is notorious in
(superconducting) samples.\cite{R33,R34,R35} Similar peaks have
already been observed in high $T_c$ compounds,\cite{R36,R37,R38}
some of them being purposely disordered.\cite{R10,R37} Therefore,
we propose to assign this peak to states localized by disorder in
our samples, whether in the chains or in the planes.

The influence of Pr ions located between the cuprate planes on the energy of
the electronic states in their vicinity has two different reasons. First, Pr
changes the energy of the carriers in each CuO$_{2}$ plane. Second, Pr
changes the electron hopping matrix element between the planes since the
hopping occurs via either Y or Pr orbitals. Due to the hopping the wave
functions of the electronic states are either even or odd with respect to
the reflection in the Y/Pr plane. The electronic density of the odd states
vanishes at the Y/Pr plane, and, therefore, Pr influences the even and the
odd states differently. Localization in the planes can be described by a
simple quantum mechanical model, a shallow 2D potential well of radius $a$
and depth $U_{0}$: 
\begin{equation}
  U(\rho )=\left\{ 
    \begin{array}{ccl}
      -U_{0} & , & \rho <a \\ 
      &  &  \\ 
      0 & , & \rho \geq a
    \end{array}
  \right. \mbox{;}
  \hspace{0.5cm} \frac{\hbar ^{2}}{m^{\ast }a^{2}}>U_{0},
  \label{eq4}
\end{equation}
where $\rho $ is the in-plane distance. Eq. \ref{eq4} implies that Pr ions
modify in their vicinity the effective crystal potential acting upon the
carriers. Therefore, $2a$ is expected to be close to the lattice parameter
in the CuO$_{2}$ plane, while $U_{0}$ depends on the parity of the state.
The depths of the well should be typically of the order of one eV due to a
relatively large difference between Pr and Y ions. However, due to
two-dimensionality of the system this potential yields a weakly bound state
with the energy $-\varepsilon _{0}\ll U_{0}$: 
\begin{equation}
  \varepsilon_0 = -\frac{2\hbar ^{2}}{m^{\ast }a^{2}}\exp 
  \Bigg(-\frac{2\hbar^{2}}{m^{\ast }a^{2}}\cdot \frac{1}{U_{0}}\Bigg),  
  \label{eq5}
\end{equation}
with the Fourier component of the wave function: 
\begin{equation}
  \psi (q)=2\sqrt{\pi }\frac{\ell }{1+\ell ^{2}q^{2}},  \label{eq6}
\end{equation}
where the radius of the bound state is $\ell =\hbar /\sqrt{2m^{\ast}
|\varepsilon _{0}|}\gg a$. An external electric field ${\bf E}={\bf E}%
_{0}e^{-i\omega t}$ (${\bf E}_{0}$ being the field amplitude) at the frequency $%
\omega \geq -\varepsilon _{0}/\hbar $ causes transitions from the localized
to band states with momentum $p=\sqrt{2m^{\ast }(\hbar \omega +\varepsilon
_{0})}.$ Using Fermi's Golden Rule, we obtain the transition rate
(probability per time unit) from the $\psi (q)$ state to the band states: 
\begin{equation}
  W(\omega )=-2\pi \hbar \frac{e^{2}}{m^{\ast }}\varepsilon _{0}
  \frac{\hbar\omega +\varepsilon _{0}}{(\hbar \omega )^{4}}
  f_{\varepsilon_{0}}(1-f_{\varepsilon _{0}+\hbar \omega })E_{0}^{2},  
  \label{eq7}
\end{equation}
where $f_{\varepsilon _{0}}$ and $f_{\varepsilon _{0}+\hbar \omega }$ are
the filling factors of the states. Eq. \ref{eq7} assumes that the
perturbation corresponding to the dipole interaction with the electric field
is $\widehat{V}=-i(e/m^{\ast }\omega )\widehat{{\bf p}}{\bf E}$, $\widehat{%
{\bf p}}$ being the momentum operator. The energy dissipation rate in a
medium (per unity volume) is proportional to $\sigma_1 (\omega )E_{0}^{2}$.
In other words, the contribution of one localized state to $\sigma_1
(\omega )E_{0}^{2}$ is proportional to the energy absorbed per transition
multiplied by the transition rate. It then follows that the optical
conductivity due to localized states is proportional to $W(\omega )$ in Eq. 
\ref{eq7} multiplied by $\hbar \omega $ and the concentration of the
localized states $N_{L}$. Neglecting the $\omega -$dependence of $f_{\varepsilon
_{0}+\hbar \omega }$, the contribution of the localized states with the
energy $\varepsilon _{0}$ to $\sigma_1 (\omega )$ at $\omega \geq -\varepsilon
_{0}/\hbar $ is then written as 
\begin{equation}
  \sigma _{L}(\omega )=-A\varepsilon _{0}\frac{\hbar \omega +
  \varepsilon _{0}}{(\hbar \omega )^{3}},  
  \label{eq8}
\end{equation}
where $A$ is a constant proportional to the concentration of the bound
states. At a temperature $T$ a part of the localized states is empty due to
thermal excitation, and the thermally activated carriers contribute in the
Drude-like way to the optical conductivity. The localized states give the
dominant contribution to $\sigma_1 (\omega \sim \left| \varepsilon _{0}\right|
/\hbar )$ if $N_{L}>N_{0}\left| \varepsilon _{0}\right| \tau /\hbar $ for $%
\left| \varepsilon _{0}\right| \tau /\hbar \gg 1$ or $N_{L}>N_{0}\hbar
/\left| \varepsilon _{0}\right| \tau $ for $\left| \varepsilon _{0}\right|
\tau /\hbar \gg 1,$ where $N_{0}$ is the concentration of the mobile
carriers participating in the Drude-like conductivity.

Virtually any conducting oxide shows a broad overdamped mid-infrared 
band (MIB). Several papers attempted to describe this band in terms 
of various excitations such as polarons,\cite{POL1,POL2,POL3} 
stripes,\cite{STRIP1} two magnon,\cite{MAG1} etc. Nevertheless, no
conclusive evidence to support any of this effects as being the MIB
exists. For any practical purposes, it is widely accepted that the
characteristics of the MIB (position and spectral weight) can be
obtained from a Lorentz oscillator. Therefore, the general case
for the conductivity composed of disordered and mid-infrared states gives 
\begin{equation}
  \sigma_1(\omega ) = \sum_{i=1}^m \sigma_L^{(i)} + 
                    \sum_{j=1}^n \frac{2\pi }{Z_0}
                    \frac{\Delta \epsilon _j \Omega _{0(j)}^2 \gamma_j\omega^2}
                         {[\Omega _{0(j)}^2-\omega^2]^2 + \gamma_j^2\omega^2},
  \label{eq9}
\end{equation}
where the first summation is on disordered states and the second is the
Lorentz MIR states.\cite{R39} $\Delta \epsilon $ is the oscillator
strength, $\Omega _{0}$ is the resonance frequency and $\gamma $ its damping.

\begin{figure}
\begin{center}
  \input epsf
  \epsfclipon
  \epsfxsize=8cm
  \epsffile{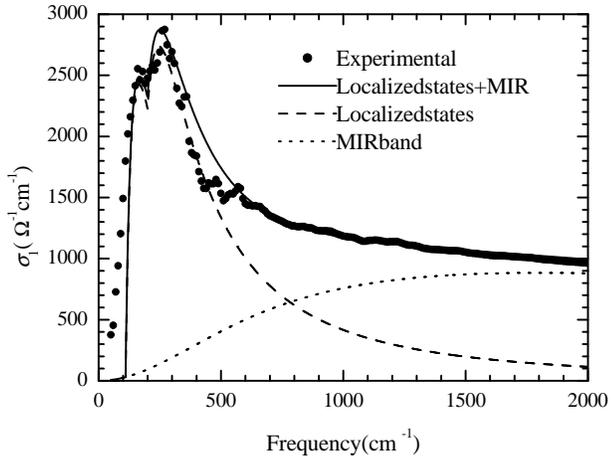}
\end{center}
\caption{IR conductivity at 50 K of the $x = 0.5$ sample. The
         experimental data are the solid circles. The solid
         line is composed of two localization peaks and a
         mid-IR (MIR) peak. The individual contributions of
         the localization peaks is show by the dashed line.
         The dotted line is the MIR contribution.}
\label{fig4}
\end{figure}

The solid circles in Fig. \ref{fig4} shows the conductivity of the $x=0.5$
sample at 50 K. Since the Pr substitution influences the odd and the even
states differently, one expects two different localization energies in the
system, which correspond to contribution of the odd and the even states in $%
\sigma _{L}(\omega )$. The solid line is a fit using the model in Eq. \ref
{eq9} with two disordered states \cite{NOTEONFIT} and one MIR oscillator.
Fitting parameters are $|\varepsilon _{0}^{(1)}|=110\mbox{ cm}^{-1}$, $%
A^{(1)}=1.8\times 10^6\Omega ^{-1}\mbox{ cm}^{-2}$, $|\varepsilon
_{0}^{(2)}|=200\mbox{ cm}^{-1}$, $A^{(2)}=1.5\times 10^6\Omega ^{-1}%
\mbox{
cm}^{-2}$, $\Delta \epsilon =90$, $\Omega _{0}=1800\mbox{ cm}^{-1}$ and $%
\gamma =5500\mbox{ cm}^{-1}$. Individual contributions from disordered
localized states and MIR bands are shown, respectively, by dashed and dotted
lines. The chosen values for $A^{(1)}$ and $A^{(2)}$ correspond to $%
N_{L}\sim 10^{14}$ cm$^{-2}$ for odd and even states and $m^{\ast }$ of the
order of the bare electron mass. Going back to Eq.~\ref{eq5} we can estimate
the radius of the potential well. Since $a\sim \sqrt{2}\hbar /\sqrt{%
U_{0}m^{\ast }\ln (U_{0}/|\varepsilon _{0}|)}$, for $\varepsilon _{0}\sim 200%
\mbox{ cm}^{-1}\mbox{ (25 meV)}$ and $U_{0}=1.0$ eV, we obtain $a$ close to $%
2$~{\AA }. This radius of the potential well indicates that the changes
really happen at the atomic level, consistent with our picture of disorder
introduced by Pr ions.

Below 700 cm$^{-1}$, the IR conductivity shown in Fig. \ref{fig4} is mostly
described by this localization model. The model reproduces the amplitude and
the asymmetric shape of the localization peak well.\cite{LOCSUPER}
The dc conductivity, however, is due to a small free carrier contribution.

Localization reconciles the apparent contradiction between the dc
resistivity and the optical conductivity in the $x = 0.5$ sample. The inset
of Fig. \ref{fig3} shows the behavior of the carrier density $N_{eff}$
(normalized here to the 300 K YBCO value). The linear parts of the
resistivity in Fig. \ref{fig1} have their slopes shown in Table \ref{tab1}. If
Matthiessen's law was roughly followed, the ratio between slopes should be
proportional to the ratio of the charge densities. The loss in spectral
weight for $x = 0.4$ follows satisfactorily the slope change observed in the
dc measurements. The ratio between the resistivity slopes is 0.79 and the
integrated charge density of the $x = 0.4$ compound equals 76 \% that from
YBCO. This suggests that what we observe here has to be mainly assigned to
actual underdoping, namely a decrease of the ratio $n/m^*$.

A striking effect appears on the $x=0.5$ Pr sample. Whereas its dc
conductivity is 5 \% of pure YBCO, the integrated carrier density is 65 \%
of that from YBCO, entirely inconsistent with the observed resistivity
increase. The slopes ratio (0.24) is not of much help in solving this
puzzle. We are thus led to conclude that the high dc resistivity in the $%
x=0.5$ sample is related to the localization of free carriers rather than to
the decrease in their concentration. The Drude-like peak related to mobile
charge carriers becomes a peak centered at 250 cm$^{-1}$ without a strong
loss of spectral weight.

Such a disorder induced localization must build up gradually with Pr
concentration and localized states have to be present in the $x=0.4$ sample
optical conductivity. The frequency dependent scattering rate of the $x=0.4$
and $x=$0.5 samples show similar structures around 250 cm$^{-1}$. At this
very frequency, the optical conductivity of the $x=0.4$ material has a
shoulder. Thus the localization peak is already present in the $x=0.4$
compound. Of course its oscillator strength is smaller and the mobile
carriers response dominates the spectra~\cite{R26} since the conditions $%
N_{L}>N_{0}\left| \varepsilon _{0}\right| \tau /\hbar $ or $N_{L}>N_{0}\hbar
/\left| \varepsilon _{0}\right| \tau $ are not yet fulfilled there.

One may remark that the loss of spectral weight in YBCO occurs at all
frequencies below $\sim 800$ cm$^{-1}$.\cite{R08} Why then is the loss of
spectral weight in the $x=0.4$ material limited to the 300--800 cm$^{-1}$
range? According to the $x=0.5$ behavior, the localization peak evolves in
the opposite direction of the normal state gap. We suggest that the
localization peak in the $x=0.4$ sample compensates the normal state gap
spectral weight loss in the 200 cm$^{-1}$ region thus explaining the
apparent discrepancy between underdoped and Pr-substituted YBCO.

\section{Summary}
In summary, the optical response of Pr-YBCO films shows the
localization of the charge carriers, thus clearly
indicating the coexistence of localization and superconductivity.
A peak in the optical conductivity due to this localization effect 
is fully resolved in the $x = 0.5$ Pr-YBCO at 250 cm$^{-1}$. 
We argue that such a localization is already present in the 
$x = 0.4$ sample and compensates the normal state gap spectral 
weight loss. The localization peak is described through in-plane 
disorder introduced by Pr atoms.

\acknowledgements
The authors are grateful to G. Deutscher and R. Combescot for
helpful comments, and to B. Briat for his
help with the visible measurements. One of us (RPSML) acknowledges
the financial support of the UE grant \# 93.2027.IL. EYS
acknowledges support from the Austrian Science Fund via the
project M591-TPH. This work was partially supported by the Oren
Family chair of Experimental Solid State Physics.


\begin{references}
\bibitem[*]{lobomail}Present address: LURE -- Univ. Paris-Sud -- 
					 BP 34 -- 91898 Orsay Cedex -- France -- 
                     e-mail: lobo@lure.u-psud.fr
\bibitem{R01} N.L. Wang, A.W. McConnell, and B.P. Clayman, Phys.
              Rev. {\bf B} 60, 14883 (1999).
\bibitem{R02} M.J. Rice, and Y.R. Wang, Phys. Rev. B {\bf 36},
              8794 (1997).
\bibitem{R03} S.Y. Tian, and Z.X. Zhao, Physica C {\bf 170}, 279
              (1990).
\bibitem{R04} C.M. Varma, P.B. Littlewood, S. Schmitt-Rink,
              E. Abrahams, and A.E. Ruckenstein, Phys. Rev. Lett. 
              {\bf 63}, 1996 (1989).
\bibitem{R05} J.C. Phillips, Phys. Rev. B {\bf 43}, 3656 (1991).
\bibitem{R06} J. Ruvalds, and A. Virosztek, Phys. Rev. B {\bf 43},
              5498 (1991).
\bibitem{R07} M. Ogata, and P.W. Anderson, Phys. Rev. Lett. {\bf 70},
              3087 (1993).
\bibitem{R10} D.N. Basov, A.V. Puchkov, R.A. Hughes, T. Strach, 
              J. Preston, T. Timusk, D.A. Bonn, R. Liang, and 
              W.N. Hardy, Phys. Rev. B {\bf 49}, 12165 (1994).
\bibitem{R13} A. Kapitulnik, and G. Kotliar, Phys. Rev. Lett {\bf 54},
              473 (1985).
\bibitem{R14} C. Huscroft, and R.T. Scalettar, Phys. Rev. Lett.
              {\bf 81}, 2775 (1998).
\bibitem{R14b} M.V. Sadovskii, Physics Reports {\bf 282}, 225 (1997).
\bibitem{R08} A.V. Puchkov, D.N. Basov, and T. Timusk, J. Phys.
              Condens. Matter {\bf 8}, 10049 (1996).
\bibitem{R09} C. Renner, B. Revaz, J.Y. Genoud, K. Kadowaki, and 
              \O. Fischer, Phys. Rev. Lett. {\bf 80}, 149 (1998).
\bibitem{R11} B. Wuyts, E. Osquiguil, M. Maenhoudt, S. Libbrecht, 
              Z. X. Gao, and Y. Bruynseraede, Phys. Rev. B {\bf 47}, 
              5512 (1993).
\bibitem{R12} T. Ito, K. Takenaka, and S. Uchida, Phys. Rev. Lett.
              {\bf 70}, 3995 (1993).
\bibitem{R12b} H. Alloul, T. Ohno, and P. Mendels, Phys. Rev.
               Lett. {\bf 63}, 1700 (1989).
\bibitem{R18} J. Fink, N. Nücker, H. Romberg, M. Alexander, 
              M.B. Maple, J. J. Neumeier, and J.W. Allen, 
              Phy. Rev. B {\bf 42}, 4823 (1990).
\bibitem{R19} A.P. Reyes, D. E. MacLaughlin, M. Takigawa, P. C. Hammel, 
              R. H. Heffner, J. D. Thompson, and J. E. Crow, Phys. Rev. B 
              {\bf 43}, 2989 (1991).
\bibitem{R20} D.P. Norton, D.H. Lowndes, B.C. Sales, J.D. Budai, 
              B.C. Chakoumakos, and H.R. Kerchner, Phys. Rev. Lett. {\bf 66},
              1537 (1991).
\bibitem{R21} A.I. Liechtenstein, and I.I. Mazin, Phys. Rev. Lett.
              {\bf 74}, 1000 (1995)
\bibitem{R22} C.H. Booth, F. Bridges, J B. Boyce, T. Claeson, Z. X. Zhao,
              and P. Cervantes , Phys. Rev. B {\bf 49}, 3432 (1994).
\bibitem{R23} A.G. Sun, L.M. Paulius, D.A. Gajewski, M.B. Maple, and 
              R.C. Dynes, Phys. Rev. B {\bf 50}, 3266 (1994).
\bibitem{R24} G.A. Levin, T. Stein, C.C. Almasan, S.H. Han, D.A. Gajewski, 
              and M.B. Maple, Phys. Rev. Lett. {\bf 80}, 841 (1998).
\bibitem{R25} M. Merz, N. Nucker, E. Pellegrin, P. Schweiss, S. Schuppler,
              M. Kielwein, M. Knupfer, M.S. Golden, J. Fink, C. T. Chen,
              V. Chakarian, Y.U. Idzerda, and A. Erb , Phys. Rev. B {\bf 55}, 
              9160 (1997).
\bibitem{R25b} R. Fehrenbacher, and T.M. Rice, Phys. Rev. Lett. 
               {\bf 70}, 3471 (1993).              
\bibitem{R15} K. Takenaka, Y. Imanaka, K. Tamasaku, T. Ito, and S. Uchida, 
              Phys. Rev. B {\bf 46}, 5833 (1992).
\bibitem{R16} H.L. Liu, M.A. Quijada, A.M. Zibold, Y.D. Yoon, D.B. Tanner, 
              G. Cao, J.E. Crow, H. Berger, G. Margaritondo, L. Forr\'o,
              O. Beom-Hoan, J.T. Markert, R.J. Kelly, and M. Onellion, 
              J. Phys. Condens. Matter {\bf 11}, 239 (1999).
\bibitem{R17} R. Buhleier, S.D. Brorson, I.E. Trofimov, J.O. White, H.U. Habermeier, 
              and J. Kuhl , Phys. Rev. B {\bf 50}, 9672 (1994).
\bibitem{R17b} C. Bernhard, T. Holden, A. Golnik, C.T. Lin, and M. Cardona,
               Phys. Rev. B {\bf 62}, 9138 (2000).  
\bibitem{R26} R.P.S.M. Lobo, N. Bontemps, D. Racah, Y. Dagan, and G. Deutscher, 
              {\it submitted}.
\bibitem{R28} J.W. Allen, and J.C. Mikkelsen, Phys. Rev. B {\bf 15},
              2953 (1976).
\bibitem{R29} For a rather comprehensive review, see A.V. Puchkov, D.N. Basov,
              and T. Timusk, J. Phys. Condens. Matter {\bf 8}, 10049 (1996).
\bibitem{NOTEONPLASMA} Written this way, $\Omega_p$ is a pulsation in MKS units
                       measured in s$^{-1}$. Using the MKS values for 
                       $\epsilon_0$, $e$, $m^*$ and the speed of light $c$ and
                       taking $n$ in cm$^{-3}$ one obtains the plasma frequency
                       in cm$^{-1}$ by 
                       $\Omega_p^2 = 10^{-4} ne^2/(2 \pi c)^2 \epsilon_0 m^*$.
\bibitem{R31} J. Sch\"utzmann,  B. Gorshunov, K. F. Renk, J. M\"unzel, A. Zibold, 
              H. P. Geserich, A. Erb, and G. M\"uller-Vogt , Phys. Rev. B {\bf 46}, 
              512 (1992).
\bibitem{R32} K. Yoshida, Phys. Rev. B {\bf 60}, 9325 (1999).
\bibitem{R33} P.W. Anderson, K.A. Muttalib, and T.V. Ramakrishnan,
              Phys. Rev. B {\bf 28}, 117 (1983).
\bibitem{R34} M. Ma, and P.A. Lee, Phys. Rev. B {\bf 32}, 5658 (1985).
\bibitem{R35} A. Kapitulnik, and G. Kotliar, Phys. Rev. Lett {\bf 54},
              473 (1985).
\bibitem{R36} A.V. Puchkov, T. Timusk, S. Doyle, and A.M. Hermann, 
              Phys. Rev. B {\bf 51}, 3312 (1995).
\bibitem{R37} D.N. Basov, B. Dabrowski, and T. Timusk, Phys. Rev. Lett.
              {\bf 81}, 2132 (1998).
\bibitem{R38} J.J. McGuire, M. Windt, T. Startseva, T. Timusk, D. Colson,
              and V. Viallet-Guillen, Phys. Rev. B {\bf 62}, 8711 (2000).
\bibitem{POL1} D.M. Eagles, R.P.S.M. Lobo, and F. Gervais, Phys. Rev. B {\bf 52}, 6440 (1995).
\bibitem{POL2} S. Lupi, P. Maselli, M. Capizzi, P. Calvani, P. Giura,
               and P. Roy, Phys. Rev. Lett. {\bf 83}, 4852 (1999).
\bibitem{POL3} T. Mertelj, D. Ku\v{s}\u{c}er, M. Kosec, and D. Mihailovic , Phys. Rev. B 
               {\bf 61}, 15102 (2000).
\bibitem{STRIP1} S. Tajima, N.L. Wang, N. Ichikawa, H. Eisaki, S. Uchida, 
                 H. Kitano, T. Hanaguri, and A. Maeda , Europhys. Lett. 
                 {\bf 47}, 715 (1999).
\bibitem{MAG1} M. Gr\"uninger, D. van der Marel, A. Damascelli, 
               A. Erb, T. Nunner, and T. Kopp, Phys. Rev. {\bf 62}, 12422 (2000).
\bibitem{R39} D.B. Tanner, and T. Timusk, in {\it Physical Properties
              of High Temperature Superconductors III}, pp.363-469,
              Ed. D.M. Ginsberg, World Scientific (1992).
\bibitem{NOTEONFIT} Using a distribution of cutoff energies does not
   improve the fit, except for rounding the minimum observed between
   the two peaks. Our best results were obtained with gaussian
   distributions having widths (FWHM) of 10 and 15 cm$^{-1}$.
   Remarking that the data resolution is 4 cm$^{-1}$ these
   distributions are basically single peaks.
\bibitem{LOCSUPER} As it was shown by W.A. Little [Phys. Rev. A {\bf 134}, 
                   1416 (1964)] polarization of localized states can lead 
                   to attraction and superconductivity of the carriers. 
                   The large polarizability in our case occurs due to the 
                   large radius $\ell $. Qualitative estimates show that 
                   at electron-electron separation less than 
                   $\ell ( l/\epsilon a_{B}) ^{1/3}$ the attraction 
                   between electrons due to the polarization of the 
                   localized states is comparable or larger than their 
                   Coulomb repulsion ($\epsilon $ is the dielectric function, 
                   and $a_{B}$ is the Bohr radius).
\end{references}
\end{document}